\documentclass[epj,referee]{svjour}
\usepackage{cite}
\usepackage{graphics}

\begin{document}
\title{Fluctuation in e-mail sizes weakens power-law correlations in e-mail flow}
\author{Yoshitsugu Matsubara\inst{1} \and Yasuhiro Hieida\inst{2} \and Shin-ichi Tadaki\inst{3}
}                     
%
%
\institute{Computer and Network Center, Saga University, Saga 840-8502, Japan\\
\email{matubara@cc.saga-u.ac.jp\inst{1}, hieida@cc.saga-u.ac.jp\inst{2}, tadaki@cc.saga-u.ac.jp\inst{3}}
}
\date{Received: date / Revised version: date}
%
\abstract{
Power-law correlations have been observed in packet flow over the Internet.
The possible origin of these correlations includes demand for Internet services.
We observe the demand for e-mail services in an organization, and analyze correlations in the flow and the sequence of send requests using a Detrended Fluctuation Analysis (DFA). 
The correlation in the flow is found to be weaker than that in the send requests.
Four types of artificial flow are constructed to investigate the effects of fluctuations in e-mail sizes.
As a result, we find that the correlation in the flow originates from that in the sequence of send requests.
The strength of the power-law correlation decreases as a function of the ratio of the standard deviation of e-mail sizes to their average.
\PACS{
      {05.40.Ca}{Fluctuation phenomena}  \and
      {05.45.Tp}{Time series analysis}  \and
      {89.20.Hh}{Internet}
     } 
} 
\maketitle

\section{Introduction} 

The Internet forms a vital communication medium in modern societies, 
and recently its structure and dynamical phenomena have attracted considerable research interest. 
For instance, the structure of the Internet has been reported to be scale-free~\cite{Faloutsos:1999},
and packet flow over the Internet has been observed to bear a power-law correlation~\cite{Paxson:1995, Csabai:1994, Takayasu:1996, Tadaki:2007}.
The mechanism of power-law correlations in the inter-event times at which e-mails are sent has also been discussed \cite{Eckmann:2004, Barabasi:2005, K.-I.:2008, Malmgren:2008, Anteneodo:2010, Karsai:2012}.

Possible origins of these power-law correlations include the structure of the Internet and the internal mechanisms and demand for Internet services. 
We focus our attention on the demand for e-mail services.

The reason for our interest in e-mail demand is as follows: users in an organization often send e-mail requests to their organizational e-mail servers. 
These servers mainly receive sending requests from their internal network. 
Therefore, the time series of these requests is not affected by the scale-free structure of the Internet.

E-mail is one of the most popular services on the Internet. 
Almost all members of an organization use e-mail services in their daily communication. 
Thus, a large number of records (logs) are available, 
and these records contain the time stamps and sizes of the e-mails.
Moreover, as they are so well used, e-mail services are provided continuously, 
meaning that sufficiently long-term records are available to analyze power-law correlations.

We analyze a three-year time series of sent e-mails. 
Two types of time series can be obtained from the records of an e-mail server.
The first concerns the flow, namely the sum of e-mail sizes for every unit time.
The number of send requests is measured also in the unit time. 
The second is the sequence of the numbers.
Using a Detrended Fluctuation Analysis (DFA), we find that power-law correlations are present in both the flow and the request sequence.

The correlation in the flow is found to be weaker than that in the requests.
We investigate the origins of the difference by analyzing artificial flows constructed with the real request sequence.
We find that fluctuations in e-mail sizes affect the correlation.

This paper is organized as follows.
In Section \nolinebreak\ref{sec2}, we briefly describe the DFA, before analyzing the power-law correlations in the flow and the requests in Section \nolinebreak\ref{sec3}.
In Section \nolinebreak\ref{sec4}, the effects of fluctuations in e-mail sizes are discussed using artificial flows.
Finally, Section \nolinebreak\ref{summary} gives a summary of our findings.

\section{Detrended Fluctuation Analysis}\label{sec2}

We employ the DFA~\cite{Peng:1994, Peng:1995}, which was developed to analyze correlations in non-stationary series, to detect power-law correlations.

The method employed in this paper can be described as follows. Consider a discrete time series $u(t) \ (0 \leq t < T)$, where $T$ is the length of $\left\{ u(t) \right\}$.
The profile $y(t)$ of $u(t)$ is defined as the accumulated deviation from the average
\begin{equation}
y(t) = \sum_{i=0}^{t}[ u(i) - \langle u \rangle ], \ (0 \leq t < T), \label{DFA_profile}
\end{equation}
where $\langle u \rangle = T^{-1}\sum_{t=0}^{T-1}u(t)$ is the average of $u(t)$.

The profile $y(t)$ is divided into $\lfloor T/l \rfloor$ non-overlapping segments of length $l$.
The local trend $\widetilde{y}_{n}(t)$ in the $n$-th segment is fitted linearly by applying the method of least squares to $y(t)$ in each segment. 
This version of the method is called the {\em first-order} DFA.
The detrended profile $y_{l}(t)$ is defined as the deviation of the profile $y(t)$ from the local trend $\widetilde{y}_{n}(t)$
\begin{equation}
y_{l}(t) = y(t) - \widetilde{y}_{n}(t), \ \mathrm{if} \ nl \leq t < (n+1)l. \label{DFA_detrended_profile}
\end{equation}

The root mean square of the deviations $F(l)$ is defined as
\begin{equation}
F(l) = \left[ \frac{1}{T} \sum_{t=0}^{T-1} y_{l}^{2}(t) \right]^{1/2}.
\end{equation}
If $F(l)$ behaves as $F(l) \sim l^\alpha$ ($1/2 < \alpha \le 1$),  the time series $u(t)$ contains power-law correlations \cite{Peng:1994}.
In this paper, the symbol ``$\sim$'' is used to denote approximate proportionality.
The exponent $\alpha$ indicates the degree of power-law correlation.
The power spectrum $P(k)$ of $u(t)$ obeys $P(k) \sim k^{- \beta}$, where $\beta = 2 \alpha - 1$. 
The value $\alpha = 1/2$ means that $u(t)$ is uncorrelated. 
A sequence with $\alpha = 1$ bears $1/f$ fluctuations.

\section{Power-law correlations in e-mail flow and send requests} \label{sec3}

We analyze the power-law correlation present in e-mail sending sequences.
The time series are obtained from our university, which has about 2,400 staff (including faculty members) and 7,300 students.
In this paper, we choose data taken from the staff e-mail server.

The e-mail server records the time stamp and the size of each e-mail for which a send request is submitted.
We focus our attentions to e-mails submitted from the inside of our university and count the number of requests which contain one or many recipients.
Thus the number of requests is equivalent to the cumulated number of senders who sent e-mails to the server.
The counting process does not distinguish the destination of each e-mail.
The temporal length of the time series is three years, from 9 May 2008 to 8 May 2011.
The series contains several defects.
These originate from weekly system maintenance, annual power-unit maintenance, and service trouble.
The effect of the weekly system maintenance can be eliminated, 
and the method of elimination is described in detail later in this section.
We assume that short-time defects have no effect on the power-law correlations that may be present.

\begin{figure}[tb]
\begin{center}
\resizebox{0.45\textwidth}{!}{\includegraphics{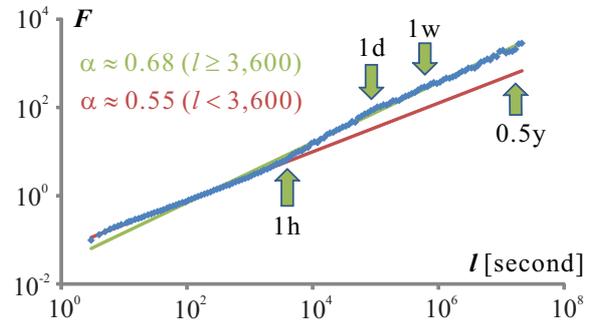}}
\caption{
The DFA result for the e-mail flow.
The temporal resolution of the e-mail flow is one second.
The horizontal axis denotes the length of segment [second] and the vertical axis denotes the root mean square deviation $F(l)$.
Four vertical arrows correspond to one hour (1h), one day (1d), one week (1w), and a half year (0.5y), respectively.
Each blue point represents the result of the DFA for each value of $l$.
One convex bending point exists at around one hour.
Two straight lines are obtained by fitting with the method of least squares.
The green line is drawn by fitting for $l \geq 3,600$ and the red line for $l < 3,600$.
These slopes correspond to the exponents of power-law correlations.
The exponent for the longer length ($l \geq 3,600$) is $\alpha \approx 0.68$, and for the shorter range ($l < 3,600$) is $\alpha \approx 0.55$. 
}\label{fig1}
\end{center}
\end{figure}

The temporal resolution of the time series is one second.
The result of applying DFA to the flow is shown in Fig. \nolinebreak\ref{fig1}, where the flow is the sum of e-mail sizes for every second.
One convex bending point exists at around one hour. 
Over the shorter time range than one hour, the exponent $\alpha \approx 0.55$, and over the longer time range $\alpha \approx 0.68$.
This crossover is a typical feature of correlated time series with uncorrelated noise in the time range shorter than the convex bending point~\cite{Tadaki:2011}. 
Such short-range noise will prevent us from analyzing long-term power-law correlations sufficiently. 
Hence, we construct a coarse-grained time series from the original data,
with the temporal resolution reduced to one hour.

The flow $f(t)$ is defined as the sum of sizes (in megabytes) of e-mails sent from $t$ to $t+1$, where $t$ is measured in hours.  
Weekly and daily periodic motions are recognized in $f(t)$, 
and these periodic motions will affect the results of the DFA~\cite{Hu:2001}. 
Thus, the modified sequence $f'(t)$ is constructed by eliminating the weekly trend:
\begin{eqnarray}
f'(t)  =  f(t) - \bar{f}_{\mathrm{week}}(t \bmod T_{\mathrm{week}}), \label{extract} \\
\bar{f}_{\mathrm{week}}(\tau_{\mathrm{w}})  =  \frac{1}{W}\sum_{w=0}^{W-1}f(w \times T_{\mathrm{week}} + \tau_{\mathrm{w}}), \label{weekly_trend}
\end{eqnarray}
where $\bar{f}_\mathrm{week}$ is the weekly trend, $0 \leq \tau_{\mathrm{w}} < T_{\mathrm{week}} \ (= 24 \times 7 \ \mathrm{hours})$, and $W$ is the number of weeks in the time series. 
By eliminating the weekly trend, any daily periodicity will, to some extent, also be removed. 

The sequence of send requests $r(t)$ is defined as the number of send events from $t$ to $t+1$, where $t$ is measured in hours.
The modified sequence $r'(t)$ is constructed by eliminating the weekly trend from $r(t)$, using the same method as for $f'(t)$.

\begin{figure}[tb]
\begin{center}
\resizebox{0.45\textwidth}{!}{\includegraphics{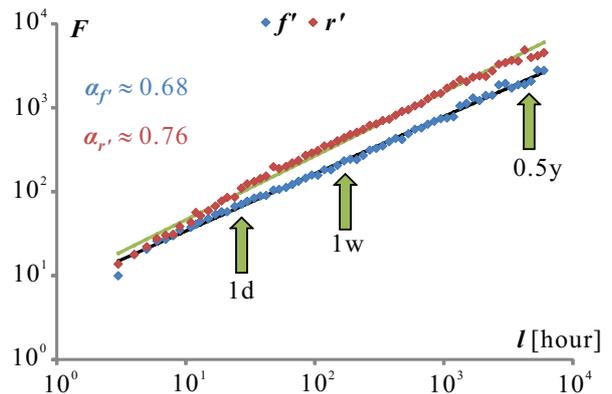}}
\end{center}
\caption{
The DFA results for the flow $f'(t)$ and the request sequence $r'(t)$.
The temporal resolution of $f'(t)$ and $r'(t)$ is one hour and the weekly trend is eliminated.
The horizontal axis denotes the length of segment [hour], and the vertical axis denotes $F(l)$.
Three vertical arrows correspond to one day (1d), one week (1w), and a half year (0.5y), respectively.
The red points denote the result of the DFA of $r'(t)$ for each value of $l$, and blue points correspond to $f'(t)$.
Two straight lines denote the best fit with the method of least squares.
The black straight line $\alpha_{f'} \approx 0.68$ fits $f'(t)$ and the green line $\alpha_{r'} \approx 0.76$ for $r'(t)$.
}\label{fig2}
\end{figure}

The DFA results for the modified flow $f'(t)$ and the modified request sequence $r'(t)$ are shown in Fig. \nolinebreak\ref{fig2}.
Using the method of least squares for all points, the exponents for $f'(t)$ and $r'(t)$ are found to be $\alpha_{f'} \approx 0.68$ and $\alpha_{r'} \approx 0.76$, respectively.
The range of the power-law correlations for $f'(t)$ and $r'(t)$ is up to half a year.
The exponent for the flow ($\alpha_{f'} \approx 0.68$) is smaller than that for the request sequence ($\alpha_{r'} \approx 0.76$).
In other words, the correlation in the flow is weaker than that in the request sequence.
The difference between the flow and the send requests may come from the fluctuation in e-mail sizes.
Thus, we focus our attention on the effects of e-mail sizes on the correlation in the flow. 

\section{Effects of fluctuations in e-mail sizes} \label{sec4}

\begin{figure}[tb]
\begin{center}
\resizebox{0.45\textwidth}{!}{\includegraphics{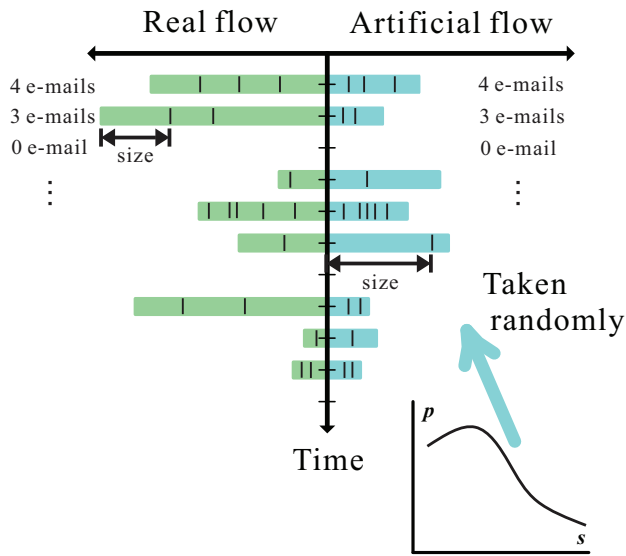}}
\end{center}
\caption{
Schematic view of constructing artificial flow.
The vertical axis denotes time, and the left- and right-hand sides of the horizontal axis denote the real and artificial flows, respectively.
Each of vertical lines in horizontal bars separates e-mails, and the length of each separated bar represents the size of the e-mail.
For each send request in the real record,  an artificial e-mail size is selected randomly from a given distribution for generating artificial flow.
}\label{fig3}
\end{figure}

As shown in the previous section, the power-law correlation in the flow is weaker than that in the send requests.
We now test whether the fluctuation in e-mail sizes may cause this difference. 
To validate this possibility, we analyze four types of artificial flow with different distributions of e-mail sizes using the real sequence of send requests.
A schematic view of constructing these artificial flows is shown in Fig. \nolinebreak\ref{fig3}.
In these artificial flows, the e-mail size for each send request is chosen randomly according to the distribution of e-mail sizes.

\subsection{Observed distribution of e-mail sizes}

\begin{figure}[tb]
\begin{center}
\resizebox{0.45\textwidth}{!}{\includegraphics{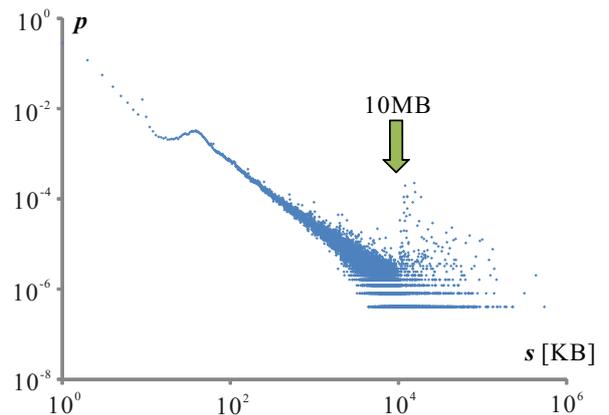}}
\end{center}
\caption{
The observed distribution $p(s)$ of e-mail sizes.
The horizontal axis denotes e-mail sizes $s$ [KB], and the vertical axis denotes the relative frequency $p(s)$ of $s$.
In the range from 100KB to 10MB, the distribution seems to obey a power law.
For reference, we obtain $p(s) \sim s^{-1.30}$ by fitting all data in $s \geq 1$KB.
The e-mail server has an upper size limit of $10 \mathrm{MB}$. 
Several e-mails exceeding this limit were repeatedly requested to be sent.
}\label{fig4}
\end{figure}

The observed distribution of e-mail size, $p(s)$, is shown in Fig. \nolinebreak\ref{fig4}, where $s$ denotes e-mail size [KB].
The average size is $\langle s \rangle \approx 230.43 \mathrm{KB}$ and the standard deviation is $\sigma \approx 1832.31 \mathrm{KB}$.
In the range from 100KB to 10MB, the distribution seems to obey a power law.
For reference, we obtain $p(s) \sim s^{-1.30}$ by fitting all data in $s \geq 1$KB.
The e-mail server has an upper size limit of $10 \mathrm{MB}$ for each send request. 
Several e-mails exceeding this limit were repeatedly requested to be sent.

\subsection{Power-law correlation in artificial flow using the observed distribution of e-mail sizes} \label{af1}

The first artificial flow $f_\mathrm{A}(t)$ is constructed as follows. For every send request in the observed time series, the size of the e-mail is randomly selected from the observed distribution $p(s)$ shown in Fig. \nolinebreak\ref{fig4}.
Then, in the same way as for the analysis of the real flow $f(t)$, we construct the time series of one-hour accumulations.
Weekly trends are also eliminated by the same method (see Section \nolinebreak\ref{sec3}).

If the sequential order of e-mail sizes for the send requests affects the power-law correlation in $f'(t)$, the correlation in the sequence of this randomized artificial flow $f_\mathrm{A}(t)$ would be different from the correlation in $f'(t)$.
The exponent ($\alpha_{\mathrm{A}} \approx 0.67$) obtained as the DFA result for the artificial time series $f_{\mathrm{A}}(t)$ is very close to that for the real flow $f'(t)$ ($\alpha_{f'} \approx 0.68$ in Fig. \nolinebreak\ref{fig2}). 
This result shows that the sequential order of e-mail size is not important to the power-law correlation in the flow.

\subsection{Power-law correlation in artificial flow using some distributions of e-mail sizes} \label{af2}

The second artificial sequence $f_\mathrm{B}(t)$ is constructed in a similar way to the first one, except for the e-mail size distribution.
We employ an exact power-law distribution $p'(s) \sim s^{-1.30}  \ \ (s \geq 1)$ with a cut-off in the size.

\begin{figure}[tb]
\begin{center}
\resizebox{0.47\textwidth}{!}{\includegraphics{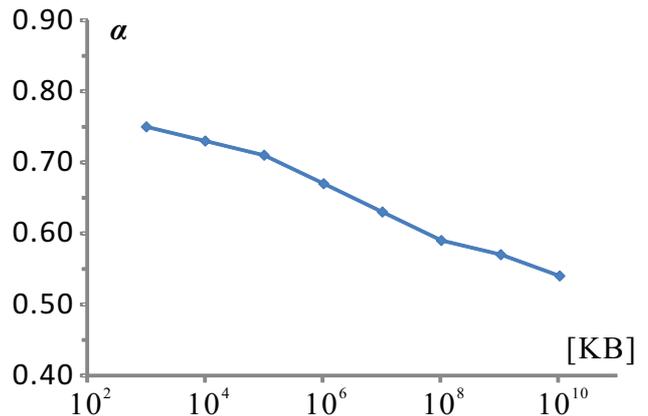}}
\end{center}
\caption{
The DFA results for artificial flow $f_{\mathrm{B}}(t)$ for each cut-off value.
The horizontal axis denotes the cut-off in e-mail size [KB], and the vertical axis denotes the exponent of the DFA result for each cut-off.
}\label{fig5}
\end{figure}

The DFA result for $f_{\mathrm{B}}(t)$ depends on the cut-off. 
The relation between the exponent $\alpha$ and the cut-off is shown in Fig. \nolinebreak\ref{fig5}. 
If the cut-off is small, $\alpha$ is close to $\alpha_{r'} \approx 0.76$ for $r'(t)$ in Fig. \nolinebreak\ref{fig2}. 
If the cut-off is large, the exponent is close to that for uncorrelated sequences ($\alpha = 1/2$). 
The result suggests that the range of fluctuation affects the power-law correlation in the flow.

We have found that the power-law correlation in e-mail flow originates from the sequence of the send requests. 
Furthermore, the wide range of fluctuations in e-mail sizes reduces the power-law correlation in the flow.

To investigate the effects of e-mail size distributions further, we construct two more sequences with different distributions.

The third artificial flow $f_\mathrm{C}(t)$ is a sequence with a uniform e-mail size distribution in $[0, M)$.
The exponent of the DFA result does not depend on the maximum $M$ of the range, and is very close to that for the sequence of the send requests ($\alpha_{r'} \approx 0.76$).

The fourth flow $f_\mathrm{D}(t)$ is constructed with an exponential distribution $p(s) \sim \mathrm{e}^{-s/s_{0}}$ for e-mail size $s$.
The exponent of the DFA result does not depend on the characteristic size $s_{0}$, and is very close to that for the sequence of send requests ($\alpha_{r'} \approx 0.76$).

\begin{figure}[tb]
\begin{center}
\resizebox{0.47\textwidth}{!}{\includegraphics{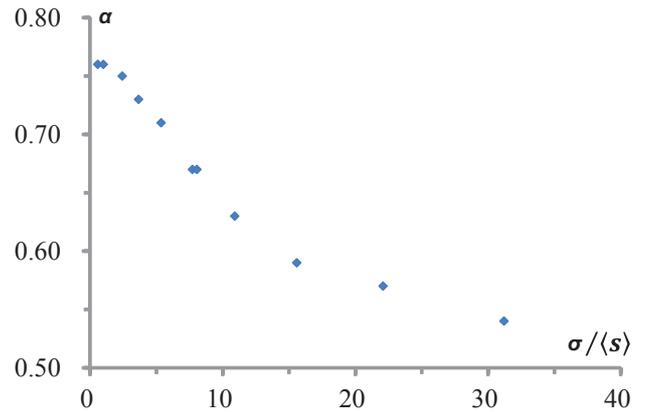}}
\end{center}
\caption{
Dependence of the DFA exponents on the ratio $\sigma/\langle s \rangle$.
The exponent decreases monotonically from $\alpha \approx 0.76$ (for the send request) to $\alpha = 1/2$ (for random noise).
}\label{fig6}
\end{figure}

The analyses in this subsection suggest that the range and standard deviation themselves do not affect the strength of the correlation in the flow.
To investigate the relation between the correlation exponents and the characteristics of the e-mail size distributions, we summarize the values of the exponents and the ratio of the standard deviation $\sigma$ to the average $\langle s \rangle$ in Table \nolinebreak\ref{table1} and Fig. \nolinebreak\ref{fig6}.
This shows that the exponent $\alpha$ decreases monotonically with the ratio $\sigma/\langle s \rangle$.

\begin{table}[tb]
\caption{
The relation between the ratio $\sigma/\langle s \rangle$ (standard deviation/average size) and the exponent $\alpha$ of the DFA result for each artificial flow.
The flow $f_{\mathrm{A}}(t)$ is the artificial flow using the real distribution of e-mail sizes (see Subsection \ref{af1}).
The flows $f_{\mathrm{B}}(t)$,  $f_{\mathrm{C}}(t)$ and $f_{\mathrm{D}}(t)$ are the artificial flows using a power-law distribution, a uniform distribution and an exponential distribution, respectively (Subsection \ref{af2}).
The exponents for $f_\mathrm{C}(t)$ and $f_\mathrm{D}(t)$ do not depend on the parameters of the distributions used for the respective flows. 
The values of the ratio $\sigma/\langle s\rangle$ for $f_\mathrm{C}(t)$ and $f_\mathrm{D}(t)$ are exact.
}
\label{table1}       
\begin{center}
\begin{tabular}{l|lrr}
\hline\noalign{\smallskip}
\multicolumn{1}{c|}{Artificial flow} && \multicolumn{1}{c}{$\sigma/\langle s \rangle$} & \multicolumn{1}{c}{$\alpha$} \\
\noalign{\smallskip}\hline\noalign{\smallskip}
$f_{\mathrm{C}}(t)$ && $1/\sqrt{3}$ & $0.76$ \\
$f_{\mathrm{D}}(t)$ && $1$ & $0.76$ \\
$f_{\mathrm{B}}(t)$ with cut-off $10^{3} \mathrm{KB}$ && $2.43$ & $0.75$ \\
$f_{\mathrm{B}}(t)$ with cut-off $10^{4} \mathrm{KB}$ && $3.66$ & $0.73$ \\
$f_{\mathrm{B}}(t)$ with cut-off $10^{5} \mathrm{KB}$ && $5.36$ & $0.71$ \\
$f_{\mathrm{B}}(t)$ with cut-off $10^{6} \mathrm{KB}$ && $7.72$ & $0.67$ \\
$f_{\mathrm{A}}(t)$ && $8.06$ & $0.67$ \\
$f_{\mathrm{B}}(t)$ with cut-off $10^{7} \mathrm{KB}$ && $10.90$ & $0.63$ \\
$f_{\mathrm{B}}(t)$ with cut-off $10^{8} \mathrm{KB}$ && $15.58$ & $0.59$ \\
$f_{\mathrm{B}}(t)$ with cut-off $10^{9} \mathrm{KB}$ && $22.09$ & $0.57$ \\
$f_{\mathrm{B}}(t)$ with cut-off $10^{10} \mathrm{KB}$ && $31.20$ & $0.54$ \\
\noalign{\smallskip}\hline
\end{tabular}
\end{center}
\end{table}

\section{Summary} \label{summary}


We used the Detrended Fluctuation Analysis (DFA) to investigate power-law correlations in the time series of the flow and send request sequence of e-mails over a three-year period.
We found that the flow and the send requests contain power-law correlations.
The exponent of the power-law correlation in the flow ($\alpha_{f'} \approx 0.68$) was observed to be smaller than that for the requests ($\alpha_{r'} \approx 0.76$).

To understand the difference between correlations in the flow and the send requests, we analyzed the correlations in artificial flows using different e-mail size distributions.
We found that the correlation in the flow originates from that in the send requests.
In addition, the exponent of the correlation in the flow was seen to decrease monotonically with the ratio of the standard deviation $\sigma$ of e-mail sizes to their average $\langle s \rangle$.
If $\sigma/\langle s\rangle$ is large, sizes of e-mails fluctuate largely.
Thus the fluctuation of e-mail sizes breaks the correlations in the sequence of send requests.
If $\sigma/\langle s\rangle$ is large enough,  the e-mail flow looks random.
On the contrary if $\sigma/\langle s\rangle$ is small, the fluctuation of e-mail sizes weakly breaks the correlations in send requests.
And the correlations in send requests are maintained if the fluctuation of e-mail sizes goes to zero.




\begin{thebibliography}{}
\bibitem{Faloutsos:1999} M. Faloutsos, M. P. Faloutsos, and C. Faloutsos. ACM SIGCOMM Comp. Commun. Rev. \textbf{29}, (1999) 251.
\bibitem{Paxson:1995} V. Paxson and S. Floyd. IEEE/ACM Trans. Networking \textbf{3}, (1995) 226.
\bibitem{Csabai:1994} I. Csabai. J. Phys. \textbf{A 27}, (1994) L417.
\bibitem{Takayasu:1996} M. Takayasu, H. Takayasu, and T. Sato. Physica \textbf{A 233}, (1996) 824.
\bibitem{Tadaki:2007} S. Tadaki. J. Phys. Soc. Jpn. \textbf{76}, (2007) 044001.
\bibitem{Eckmann:2004} J.-P. Eckmann, E. Moses and D. Sergi. Proc. Natl. Acad. Sci. USA \textbf{101}, (2004) 14333.
\bibitem{Barabasi:2005} A.-L. Barab\'{a}si. Nature \textbf{435}, (2005) 207.
\bibitem{K.-I.:2008} K.-I. Goh and A.-L. Barab\'{a}si. EPL \textbf{81}, (2008) 48002. 
\bibitem{Malmgren:2008} R.D. Malmgren, D.B. Stouffer, A.E. Motter, and L.A.N. Amaral. Proc. Natl. Acad. Sci. USA \textbf{105}, (2008) 18153. 
\bibitem{Anteneodo:2010} C. Anteneodo, R.D. Malmgren, and D.R. Chialvo. Eur. Phys. J. B \textbf{75}, (2010) 389. 
\bibitem{Karsai:2012} M. Karsai, K. Kaski, A.-L. Barab\'{a}si, J. Kert\'{e}sz. Scientific Reports. \textbf{2}, (2012) 397.
\bibitem{Peng:1994} C.-K. Peng, S. V. Buldyrev, S. Havlin, M. Simons, H. E. Stanley, and A. L. Goldberger. Phys. Rev. \textbf{E 49}, (1994) 1685.
\bibitem{Peng:1995} C.-K. Peng, S. Havlin, H. E. Stanley, and A. L. Goldberger. Chaos \textbf{5}, (1995) 82.
\bibitem{Tadaki:2011} S. Tadaki. Comp. Phys. Commun. \textbf{182}, (2011) 237.
\bibitem{Hu:2001} K. Hu, P. Ch. Ivanov, Z. Chen, P. Carpena and H. E. Stanley. Phys. Rev. \textbf{E 64}, (2001) 011114.
\end{thebibliography}
\end{document}